\documentclass [12pt]{article}
\usepackage {a4wide}
\usepackage {graphicx}
\usepackage{amsmath}

\newcommand{\dd}{{\text d}}

\title{\bf AN APPROACH TO RELATE THE WEAK AND GRAVITATIONAL INTERACTIONS}
\author{J\'ulius Vanko$^1$ and Jozef \v{S}ima$^2$ and Miroslav S\'{u}ken\'{\i}k$^2$ \\[1ex]
  $^1$Comenius University, Mlynská dolina F1, 842 48 Bratislava, Slovakia
  \\
  $^2$Slovak Technical University, FCHPT, Radlinsk\'{e}ho 9, \\
  812 37 Bratislava, Slovakia
  \\
  e-mail:  vanko@fmph.uniba.sk, \\jozef.sima@stuba.sk, sukenik.miroslav@stonline.sk}
 \date{}
\begin{document}
\maketitle
\begin{abstract}
Stemming from simple postulates of 
nondecelerative nature of the universe expansion and Vaidya metric 
application, the paper offers some dependences and relations between the 
gravity and weak interactions. It presents a mode of independent 
determination of the mass of vector bosons Z and W, and it derives the time 
of separation of electromagnetic and weak interactions. Comparisons of 
theoretically derived and experimentally obtained data indicate the 
relevancy of the used mode and provide a hint for further investigation. The 
W and Z bosons mass of about 100 GeV, together with the time and the 
Universe radius of electromagnetic and weak interactions separation 
approaching $t_{x}$ = 10$^{ - 10}$ s and $a_{x}$ = 10$^{ - 2}$ m, 
respectively, were obtained using our approach. The above values match well 
those commonly accepted. 
\end{abstract}
\textit{Keywords: Weak interactions; Gravitational interactions; Boson mass; }

\noindent
\textit{PACS (2006): 12.10.Kt, 12.20.-m, 98.80.-k}

\section{Introduction}  

The issue of unification of all four known physical forces is still one of 
the top-ten evergreens of theoretical physics. There are several approaches 
allowing to come to theoretical results which, in principle, do not 
contradict to the generally accepted physical principles and offer at least 
a partial solution of the matter. Among the approaches, the Standard Model 
[1--3] (which does not, however, provide a complete description of Nature 
since it does not address gravitational interactions), and the Superstring 
Theory [4--6], trying to become the Theory of Everything are to be mentioned. 
Within the elaboration of the mentioned and other theoretical approaches and 
their verification based on available experimental cosmological and particle 
physics data, several important achievements have been reached. 

Frequently, the approaches involve the inflationary Universe [7] as a 
starting point. In this approach, the observable part of the Universe (its 
radius $a)$ is emerging by the velocity of the light and, in turn, the Universe 
mass is gradually increasing. The same results may be obtained applying a 
hypothesis on nondecelerative expansion of the Universe (without the 
previous inflationary phase) with mass creation [8, 9]. The approaches 
differ mainly in metrics used since to describe the mass creation, the 
Vaidya metric [10, 11] or another metric implicitly involving such a 
creation must be involved. The Vaidya metric and the hypothesis on the 
matter creation has manifested its justification when explained both some 
macroworld (cosmological) and microworld (particle physics) issues, such as 
rationalization and prediction of neutron star properties [8], questions 
concerning the entropy of the universe [9], estimation of lower and upper 
mass limits of black holes [12], explanation of Podkletnov's phenomenon 
[13], clarification of the internal structure and parameters of the hydrogen 
atom [14], prediction of bands in far-infrared low-temperature spectra of 
chemical compounds [15]. 

A possibility of our approach to contribute to understanding the 
interactions of gravitational, electrostatic and electromagnetic fields [16] 
we have taken as a challenge to exploit the approach in an attempt to offer 
the applicability of Vaidya metric and the approach of nondecelerative 
expansion of the Universe in understanding of weak and gravitational forces 
unification. The results obtained are presented in this paper.

\section{Energy of Z and W bosons }

In the early stage of the Universe creation, i.e. in the leptons era an 
equilibrium of protons and neutrons formation existed at the temperature 
about 10$^{9}$ - 10$^{10}$ K. The amount of neutrons was stabilized due to 
weak interactions, which were responsible for processes such as 
\begin{equation}
\label{eq1}
\tilde {\nu } + p^ + \to n + e^ + 
\end{equation}
\begin{equation}
\label{eq2}
e^ - + p^ + \to n + \nu 
\end{equation}
The cross section $\sigma $ corresponding to the above processes may be 
expressed [17, 18] as
\begin{equation}
\label{eq3}
\sigma \cong \frac{g_F^2 \;E_w^2 }{(\hbar \,c)^4}
\end{equation}
where $g_{F}$ is the Fermi constant (1.4 $\times $ 10$^{ - 62}$ J m$^{3})$, 
$E_{w}$ is the energy of weak interactions that, based on (\ref{eq3}), can be 
formulated by relation
\begin{equation}
\label{eq4}
E_w \cong \frac{r\,\hbar ^2\,c^2}{g_F }
\end{equation}
where $r$ represents the effective range of weak interactions. Stemming from 
relation (\ref{eq4}) it holds that in a limiting case when
\begin{equation}
\label{eq5}
r = \frac{\hbar }{m_W \,c}
\end{equation}
the maximum energy of weak interaction is given by
\begin{equation}
\label{eq6}
E_w \cong m_W \,c^2
\end{equation}
Relations (\ref{eq5}) and (\ref{eq6}) represent the Compton wavelength of the vector bosons 
Z and W, and their energy, respectively. Equations (\ref{eq4}), (\ref{eq5}) and (\ref{eq6}) lead to 
the following expression for the mass of the bosons Z and W (indicated 
further on $m_{W})$
\begin{equation}
\label{eq7}
m_W^2 \cong \frac{\hbar ^3}{g_F \,c} \cong \left| {100\;{\rm GeV}} 
\right|^2
\end{equation}
providing the value that is in good agreement with the currently accepted 
value [19].

\section{Localization of gravitational energy}

The following part deals with a mode of expressing the localization of 
gravitational energy based on ideas of nondecelerative expansion of the 
Universe and application of Vaidya metric. Divergence of Einstein relation
\begin{equation}
\label{eq8}
R_{ik} - \frac{1}{2}g_{ik} \,R = \frac{8\,\pi \,G}{c^4}T_{ik} 
\end{equation}
leads, in case of weak fields, to 
\begin{equation}
\label{eq9}
\varepsilon = - \frac{R\,c^4}{8\,\pi \,G}
\end{equation}
Inside the body, the scalar curvature $R$ can be obtained by calculation. It 
follows that 
\begin{equation}
\label{eq10}
\varepsilon = \varepsilon _s 
\end{equation}
where $\varepsilon _s $ is the energy density of source. 

For the region outside the body, 
\begin{equation}
\label{eq11}
\varepsilon = \varepsilon _g 
\end{equation}
where $\varepsilon _g $ is the energy density of gravitational field. 

Scalar curvature is calculable only in the case of Vaidya metric [10, 11] 
application. The necessity of the Vaidya metric introduction deserves some 
words of justification. Suppose, the Universe horison is expanding by the 
velocity of light $c$, i.e.
\begin{equation}
\label{eq:12a}
\dd a=c  \dd t  
\end{equation}
and
\begin{equation}
\label{eq:13a}
a = c t_U
\end{equation}
where $a$ is the radius of the visible part of the Universe, $t_{U}$ is the 
cosmological time. At the same time, new mass is emerging at the horison, 
i.e. the mass of the visible part of the Universe $m_{U}$ is increasing 
obeying the relation
\begin{equation}
\label{eq12}
\frac{{\rm d}m_{\rm U} }{{\rm d}t} = \frac{m_{\rm U} }{t_{\rm U} }
\end{equation}
Applying Vaidya metric and using relations (12), (13) and (\ref{eq12}), the scalar 
curvature $R$ outside the body is obtained in the form
\begin{equation}
\label{eq13}
R = \frac{3\,R_g }{a\,r^2}
\end{equation}
where $R_g $ is the gravitational radius of a body with the mass $m$, $r$ is the 
distance. Inserting (\ref{eq13}) into (\ref{eq9}), a formula for gravitational energy 
density $\varepsilon _g $ is obtained
\begin{equation}
\label{eq14}
\varepsilon _g = - \frac{R\,c^4}{8\,\pi \,G} = - \frac{3\,m\,c^2}{4\,\pi 
\,a\,r^2}
\end{equation}

\section{Relation between the weak and gravitational interactions}

The Universe radius $a$ reaches at present
\begin{equation}
\label{eq15}
a \cong 1.3\times 10^{26}\;{\rm m}
\end{equation}

As a starting point for unifying the gravitational and weak interactions, 
the conditions in which the weak interaction energy $E_{w}$ and the 
gravitational energy $E_{g}$ of a hypothetic body with a limit mass 
$m_{lim}$
\begin{equation}
\label{eq16}
E_w = \left| {E_g } \right|
\end{equation}
can be chosen. Based on relations (\ref{eq4}) and (\ref{eq14}) in such a case it holds
\begin{equation}
\label{eq17}
\frac{r\,\hbar ^2\,c^2}{g_F } = \left| {\int {\varepsilon _g {\rm d}V} } 
\right| = \frac{m_{\lim } \,c^2\,r}{a}
\end{equation}
where $r$ is the effective range of weak interaction. It follows from (\ref{eq17}) that
\begin{equation}
\label{eq18}
m_{\lim } \cong \frac{a\,\hbar ^2}{g_F }
\end{equation}
The above relation manifests that the limit mass depends on the Universe 
radius, i.e. it is increasing with time. 

Let us see what happens if the limit mass equals the Planck mass $m_{Pc}$ 
(2.1767 $\times $ 10$^{ - 8}$ kg)
\begin{equation}
\label{eq19}
m_{\lim } = m_{Pc} 
\end{equation}
It stems from (\ref{eq18}) and (\ref{eq19}) that
\begin{equation}
\label{eq20}
a_x \cong \frac{m_{Pc} \,g_F }{\hbar ^2} \cong 10^{ - 2}\;{\rm m}
\end{equation}
and
\begin{equation}
\label{eq21}
t_x \cong 10^{ - 10}\;{\rm s}
\end{equation}
This is actually the time when, in accordance with the current knowledge, 
electromagnetic and weak interactions separated. In the time $t_{x}$ it had 
to hold
\begin{equation}
\label{eq22}
\frac{m_{Pc} }{m_W } = \left( {\frac{a_x }{l_{Pc} }} \right)^{1 / 2}
\end{equation}
Substitution of (\ref{eq20}) into (\ref{eq22}) leads to (\ref{eq7}) which means that the mass of the 
vector bosons Z and W as well as the time of separation of the 
electromagnetic and weak interactions are directly obtained, based on the 
used approach, in an independent way.

If Planck length $l_{Pc} $ is substituted for $a$ in equation (\ref{eq18}), the limit 
mass will approach to 10$^{ - 41}$ kg corresponding to the rest energy of 
10$^{ - 5}$ eV. It might represent a rest energy of some of the neutrinos.

\section{Conclusions}

\begin{enumerate}
\item The Vaidya metric allowing to localize the gravitational energy exhibits 
its capability to manifest some common features of the gravitational and 
weak interactions.

\item The paper presents an independent mode of determination of the mass of 
vector bosons Z and W, as well as the time of separation of the 
electromagnetic and weak interactions. The mode follows directly from the 
ability to localize gravitational energy density ouside a body.

\item The paper follows up our previous contributions showing the unity of the 
fundamental physical interactions. It might suggest the existence of a 
deeper relation of the weak and gravitational interactions and a common 
nature of the both interactions before thein separation. The paper can be 
considered as a hint for verification and justification of the chosen 
procedure, introduction of Vaidya metric in particular. 

\end{enumerate}

\subsection{Acknowledgments}

The financial support by the Slovak grant agency VEGA (Project No. 
1/8315/01) is gratefully acknowledged.

\section*{References}
\begin{description}

\item [][1] F. G\"{u}rsey, P. Ramond, and P. Sikivie: ``A universal gauge theory 
model based on E$_{6}$'', \textit{Phys. Lett. B, }Vol. 60, (1976), pp. 177-180.

\item [][2] W.N. Cottingham and D.A. Greenwood: ``\textit{An Introduction to the Standard Model of Particle Physics}'', Cambridge University Press, 
Cambridge, 1998, pp.256.

\item [][3] R. Oerter: ``\textit{The Theory of Almost Everything: The Standard Model, the Unsung Triumph of Modern Physics}'', Pi Press, 2005, pp.336.

\item [][4] D.J. Gross, J.A. Harvey, E.J. Martinec, and R. Rohm: ``Heterotic 
Strings'', \textit{Phys. Rev. Lett., }Vol. 54, (1985), 502-505.

\item [][5] \textit{Superstrings: The First Fifteen Years}, (Ed. J.H. Schwartz), World Scientific, Singapore, 1985.

\item [][6] M.B. Green: ``\textit{The Elegant Universe: Superstrings, Hidden Dimensions, and the Quest for the Ultimate Theory}'', W.W. Norton and Company, New York, 1999

\item [][7] A.H. Guth: ``The Inflationary Universe: The Quest for a~New Theory of 
Cosmic Origin'', Perseus Books Group, New York, 1998 

\item [][8] J. \v{S}ima and M. S\'{u}ken\'{\i}k: ``Neutron Stars - Rationalization and Prediction 
of Their Properties by the Model of Expansive Nondecelerative Universe'', in 
\textit{Progress in Neutron Star Research }(A.P. Wass, ed.), Nova Science Publishers, New York, 2005.

\item [][9] J. \v{S}ima and M. S\'{u}ken\'{\i}k: ``Entropy -- Some Cosmological Questions 
Answered by Model of Expansive Nondecelerative Universe'', \textit{Entropy, }Vol. 4, (2002), 
pp. 152-163. 

\item [][10] P.C. Vaidya: ``The Gravitational Field of a Radiating Star'', \textit{Proc. Indian Acad. Sci}. A, Vol. 
33, (1951), pp. 264-276. 

\item [][11] K.S. Virbhadra: ``Energy and Momentum in vaidya Spacetime'', \textit{Pramana -- J. Phys., }Vol. 38, 
(1992), pp. 31-35.

\item [][12] J. \v{S}ima and M. S\'{u}ken\'{\i}k: ``Black Holes - Estimation of Their Lower and 
Upper Mass Limits Stemming from the Model of Expansive Nondecelerative 
Universe'', \textit{Spacetime and Substance, }Vol. 2$, $(2001), pp. 79-81.

\item [][13] M. S\'{u}ken\'{\i}k and J. \v{S}ima: ``Podkletnov's Phenomenon - Gravity Enhacement 
or Cessation?'', \textit{Spacetime and Substance, }Vol. 2, (2001), pp. 125-129. 

\item [][14] J. \v{S}ima and M. S\'{u}ken\'{\i}k: ``The Hydrogen Atom - A Common Point of 
Particle Physics, Cosmology, and Chemistry'', \textit{Spacetime and Substance, }Vol. 3, (2002), pp. 31-34. 

\item [][15] J. \v{S}ima and M. S\'{u}ken\'{\i}k: ``Far-infrared low-temperature spectra of 
chemical compounds -- gravitational effects'', \textit{Spacetime and Substance, }Vol. 6$, $(2005), pp. 49-52.

\item [][16] J. \v{S}ima and M. S\'{u}ken\'{\i}k: ``Interaction of Gravitational, Electrostatic 
and Electromagnetic Fields -- Its Impact on Physical Phenomena and Modes of 
Experimental Verification'', \textit{Spacetime and Substance, }Vol. 4, (2003), pp. 169-173. 

\item [][17] I. L. Rozentahl: ``Physical Laws and Numerical Values of the 
Fundamental Constants'', \textit{Adv. Math. Phys. Astr.}, Vol. 31, (1986), 241-259

\item [][18] L.B. Okun, \textit{Leptons and Quarks}, Nauka, Moscow, 1981

\item [][19] S. Eidelman, K.G. Hayes, K.A. Olive, M. Aguilar-Benitez, C. Amsler, D. 
Asner, K.S. Babu, et al.$, $Review of Particle Physics, \textit{Phys. Lett. B, }Vol. 592, (2005), pp. 
335-404 
\end{description}

\end{document}